\documentclass[twocolumn, aps, prd]{revtex4}

\usepackage{amsmath,amsthm,latexsym,amssymb,amsfonts}
\usepackage{graphicx}
\usepackage{subfigure}

\DeclareMathOperator{\sgn}{sgn}

\begin{document}

\title{Singularity avoidance in the hybrid quantization of the Gowdy model}

\author{Paula Tarr\'{i}o}
\email{paula@grpss.ssr.upm.es}
\affiliation{Escuela T\'ecnica Superior de Ingenieros de Telecomunicaci\'on, Universidad Polit\'ecnica de Madrid, 28040 Madrid, Spain}
\author{Mikel Fern\'andez-M\'endez}
\email{m.fernandez.m@csic.es}
\affiliation{Instituto de Estructura de la Materia, IEM-CSIC, Serrano 121, 28006 Madrid, Spain}
\author{Guillermo A. Mena Marug\'an}
\email{mena@iem.cfmac.csic.es}
\affiliation{Instituto de Estructura de la Materia, IEM-CSIC, Serrano 121, 28006 Madrid, Spain}

\begin{abstract}
One of the most remarkable phenomena in Loop Quantum Cosmology is that, at least for homogeneous cosmological models, the Big Bang is replaced with a Big Bounce that connects our universe with a previous branch without passing through a cosmological singularity. The goal of this work is to study the existence of singularities in Loop Quantum Cosmology including inhomogeneities and check whether the behavior obtained in the purely homogeneous setting continues to be valid. With this aim, we focus our attention on the three-torus Gowdy cosmologies with linearly polarized gravitational waves and use effective dynamics to carry out the analysis. For this model, we prove that all the potential cosmological singularities are avoided, generalizing the results about resolution of singularities to this scenario with inhomogeneities. We also demonstrate that, if a bounce in the (Bianchi background) volume occurs, the inhomogeneities increase the value of this volume at the bounce with respect to its counterpart in the homogeneous case.
\end{abstract}

\maketitle

\section{Introduction}

Under quite general conditions, Einstein's general relativity predicts the appearance of spacetime singularities, such as the cosmological Big Bang, where the matter density and the curvature become infinite and therefore the theory breaks down \cite{HP,Wald}. Candidate theories for quantum gravity (e.g., string theory, M-theory, Loop Quantum Gravity, etc.) try to overcome this problem by unifying general relativity with quantum physics. Clearly, apart from solving singularities, the quantum effects introduced in these theories should be compatible with the classical behavior of gravity at large scales in spacetime regions like the one we observe.

Loop Quantum Gravity (LQG) is one of the most promising candidates to face these challenges \cite{LQG1,LQG2,LQG3}. It is based on a nonperturbative and background-independent approach which describes the geometry of the universe within the paradigm of quantum physics. The application of LQG to cosmology is known as Loop Quantum Cosmology (LQC) \cite{bojowald2011quantum,LQC1,LQC2,mena2011loop} and, to the present day, has succeeded in solving many of the problems related with spacetime singularities in a variety of \emph{homogeneous} cosmological models. In particular, the loop quantization program has been applied in detail to homogeneous and isotropic Friedmann-Robertson-Walker (FRW) models (flat \cite{APS1,APS2,ashtekar2006quantum,ashtekar2008robustness}, closed \cite{Closed,ashtekar2007loop}, and open cases \cite{vandersloot2006loop}, with and without cosmological constant \cite{bentivegna2008anti,kaminski2010loop}), and also to some homogeneous but anisotropic models (Bianchi~I~\cite{chiou1,chiou2,martin2008loop, martin2009physical, ashtekar2009loopBianchiI, szulc2008loop}, Bianchi II \cite{ashtekar2009loopBianchiII}, and Bianchi IX \cite{wilson2010loop}). All the analyses carried out so far in homogeneous scenarios indicate that (strong) curvature singularities are resolved. For example, the Big Bang and Big Crunch singularities are replaced with a Big Bounce, which connects the initially considered branch of the universe with another branch where Einstein's equations are also valid once the bouncing regime is over \cite{APS1,APS2,ashtekar2006quantum}. As a result, far from the Planck regime, the results of LQC match those of general relativity. An intuitive way to understand this is thinking that LQC corrections create a repulsive force that dominates at Planck scale but quickly vanishes at larger scales. In this way, LQC meets both the ultraviolet (small scale) and infrared (large scale) challenges.

LQC has recently been employed in order to face the quantization of \emph{inhomogeneous} spacetimes. In these cases, the strategy has been the following. After splitting the system into a homogeneous and an inhomogeneous part, a hybrid quantization approach is adopted, which combines a loop quantization for the homogeneous part~\cite{LQC1} and a Fock quantization for the inhomogeneities~\cite{QFT1}. This approach has been applied to the vacuum Gowdy model~\cite{gowdy1974models1,gowdy1974models2} (with linearly polarized gravitational waves and three-torus spatial topology), confirming that the quantum analog of the cosmological singularity is avoided~\cite{mercehybrid,mena2009hybrid,garay2010inhomogeneous}. The model has recently been revisited to include a scalar field as matter content~\cite{gowdymatters}.

Let us review very briefly the fundamentals of LQC and how it has been used in cosmological models like those mentioned above (for more details see, e.g., Refs.~\cite{ashtekar2010big,mena2011loop}). In LQG, the phase space is described by the Ashtekar-Barbero variables: an $SU(2)$ connection $A^i_a$ and its canonically conjugate momentum, the densitized triad $E_i^a$. The holonomies of the connection and the fluxes of the triads through surfaces form an algebra under Poisson brackets, which is chosen as the fundamental algebra of phase space variables. The quantization of this phase space is not carried out following standard methods (like, e.g., a Schr\"{o}dinger quantization). Instead, a different approach known as loop or polymer quantization is employed \cite{LQG1,LQG2,LQG3}. The choice of this quantization is dictated by the requirements of background independence and symmetry invariance (diffeomorphism invariance in full LQG).

The FRW models, which correspond to homogeneous and isotropic universes, are the simplest cosmological models.
In this case, the geometry of the universe evolves according to a single scale factor $a(t)$. Consequently, in LQC, the corresponding holonomies of the connection and the fluxes of the triad are determined, respectively, just by one homogeneous variable. We call $c$ and $p$ the resulting pair of variables, which are chosen to be canonical up to a factor: $\{c,p\}=\kappa/3$, where $\kappa=8\pi G\gamma$ ($G$ being Newton's constant and $\gamma=0.2375$ being the Immirzi parameter \cite{LQC1,LQC2}). It is common to use the orthonormal eigenbasis of the flux (triad) variable $p$ to represent the theory. This basis has the form $\lbrace | p\rangle: p \in \mathbb{R},\langle \tilde{p} |  p\rangle =\delta_p^{\tilde{p}} \rbrace$, so that the spectrum of $p$ is discrete but runs over the whole real line. The emergence of the discrete inner product in this $p$-basis is a characteristic of the polymer quantization \cite{LQC1,LQC2}. In this polymer representation, the explicit action of the flux and holonomy operators is given by: $\hat{p}\left| p\right\rangle = \frac{\kappa}{6}  p\left| p\right\rangle$ and $\hat{N}_\mu \left| p\right\rangle =\left| p+\mu\right\rangle$. Here, $N_\mu=\exp(i\mu c/2)$, and the holonomy elements are linear combinations of these exponential functions, with $\mu$ any real number. With this representation, the connection is not continuous; hence, indeed there is no operator representing $c$. Classically, curvature can be expressed as a limit of holonomies around a loop as the area enclosed by it shrinks to zero. However, in LQG, the geometric area has a discrete spectrum. If we call $\Delta$ the minimum of the allowed nonzero eigenvalues \cite{LQG1,LQG2,LQG3}, only areas larger than $\Delta$ are considered, and the limit of vanishing area is not reached. Instead, the usual procedure is to choose the length $\bar{\mu}$ in such a way that the physical area coincides with the minimum allowed value, i.e., with $\Delta$. This requirement can be argued to lead to the relation $\bar{\mu} = \sqrt{\Delta/|p|}$. The inclusion of this choice is usually called improved dynamics \cite{ashtekar2006quantum} in the LQC literature.

In extending LQC to more general situations, the next step consists in including anisotropies. One of the simplest anisotropic (homogeneous) cosmologies is the Bianchi I model, which describes a universe with three different scale factors (one for each direction). The phase space of this model may be seen as three copies of the FRW model. The metric for the (orthogonal) Bianchi I spacetime can be written:
\begin{equation}\label{metricBI}
ds^2 = -N^2 dt^2 + a_1^2dx^2 + a_2^2dy^2 + a_3^2dz^2,
\end{equation}
where $N$ is the lapse function and $a_i$ are the directional scale factors ($i=1,2,3$). The Ashtekar-Barbero variables for LQC are the connection functions $c_i$ and the triad functions $p_i$, whose nonvanishing Poisson brackets are:
\begin{equation}\label{eq:poissonbrackets}
\left\{c_i, p_j \right\}=\kappa\delta_{ij}.
\end{equation}
We note that there is a factor of 3 difference in these brackets with respect to the homogeneous case, owing to the different number of dimensions of the respective phase spaces.

The triad functions are related with the scale factors by $\left|p_i\right| = l_j l_k a_j a_k$ ($i\neq j \neq k $), where the $l_i$'s are fiducial lengths that, for compact topologies, can be chosen equal to the natural period $2\pi$ of each of the spatial coordinates. Besides, under quantization, one can restrict all considerations, e.g., to the sector with positive orientations of the triad components without loss of generality \cite{mena2009hybrid,garay2010inhomogeneous}, thus making the absolute value disappear in the above formula. In this case,
\begin{equation}\label{eq:p_i}
p_1 = 4 \pi^2 a_2 a_3, \ \ p_2 = 4 \pi^2 a_1 a_3, \ \ p_3 = 4 \pi^2 a_1 a_2 .
\end{equation}
In the corresponding improved dynamics scheme, the lengths of the edges of the closed loop over which holonomies are evaluated are proportional to:
\begin{equation}\label{mui}
\bar{\mu}_1=\sqrt{\Delta \frac{p_1}{p_2 p_3}}, \ \ \bar{\mu}_2=\sqrt{\Delta \frac{p_2}{p_1 p_3}}, \ \ \bar{\mu}_3=\sqrt{\Delta \frac{p_3}{p_1 p_2}}.
\end{equation}

One step further in the generalization of LQC involves the inclusion of inhomogeneities. The simplest inhomogeneous cosmologies are the Gowdy models, which have two spatial Killing vector fields and spatial sections of compact topology~\cite{gowdy1974models1,gowdy1974models2}. Among them, in this work we will consider the simplest case, namely, the model with the spatial topology of a three-torus and linearly polarized gravitational waves. To carry out the hybrid quantization program, we first split the phase space of the system into a homogeneous sector and an inhomogeneous one.

The homogeneous sector corresponds to a vacuum Bianchi I model with three-torus topology, to which the considered Gowdy model reduces when all the inhomogeneities vanish. It is represented by the connection and triad coefficients $c_i$ and $p_i$ as before, where now $i = \theta, \sigma, \delta$ are the three spatial coordinates [playing the same role as $x,y,z$ in Eq. (\ref{metricBI}); our new notation emphasizes that they are now axial coordinates]. In particular, $\sigma$ and $\delta$ are coordinates adapted to the Killing fields. We can add as well a homogeneous massless scalar field, described by the canonical pair $(\phi, p_\phi)$, with Poisson brackets $\lbrace \phi, p_\phi\rbrace=1$. On the other hand, the inhomogeneous sector is characterized by the metric field $\xi(\theta)$ and its conjugate momentum $P_{\xi}(\theta)$. The field $\xi(\theta)$ can be interpreted as the logarithmic norm of one of the Killing vectors of the model, suitably scaled and with the zero mode removed \cite{mena2009hybrid} . The canonical pair can be decomposed into nonzero Fourier modes, obtaining an infinite countable set of canonical pairs, $\lbrace (\xi_m, P_{\xi}^m): m \in \mathbb{Z} - \lbrace 0 \rbrace \rbrace$. For these modes, annihilation and creation variables $(a_m, a^*_m)$ can be introduced as follows, neglecting in principle any mass term. We define
\begin{equation}\label{eq:creationvariables}
a_m=\sqrt{\frac{\pi}{8G|m|}}\left(|m|\xi_m + i \frac{4G}{\pi}P_{\xi}^m\right),
\end{equation}
together with its complex conjugate $a^*_m$. The associated Poisson brackets are $\lbrace a_m, a^*_{\tilde{m}}\rbrace  =- i\delta_{m\tilde{m}}$. Let us also comment that, if a minimally coupled scalar field with the same symmetries of the metric is introduced in the model, the matter inhomogeneities (after an adequate scaling) can be described exactly in the same manner as the gravitational waves \cite{gowdymatters}.

The combination of the two sectors is a nontrivial process, inasmuch as they get coupled by the Hamiltonian constraint of the system, which relates the variables that describe the homogeneous degrees of freedom with the inhomogeneous variables. As mentioned above, the quantization of this Gowdy model, both with and without matter, has recently been obtained using this hybrid approach, based on the assumption that the most relevant quantum geometry effects are those that affect the homogeneous sector representing the Bianchi-I background, on which the gravitational waves propagate \cite{garay2010inhomogeneous,gowdymatters}.

Since a complete quantum description of this model (and similar ones) may be too complicated to handle, considerable attention has been paid to what is claimed to be the effective dynamics of the model, assumed to describe the evolution of the peaks of appropriate semiclassical states in regimes governed by classical equations corrected with quantum effects \cite{effective,effectsingh}. For the homogeneous and isotropic case, this effective dynamics has been deduced analytically under certain conditions on the considered semiclassical states \cite{effective}. Besides, all numerical simulations in anisotropic models support the use of this deduced effective dynamics, when straightforwardly extended to the anisotropic scenarios \cite{chioueffective}. In this context, it seems natural to apply the same philosophy to the Gowdy model and extend to it the considered effective dynamics, at least as a way to explore the possible implications and self-consistency of the approach. The model that we consider is parametrized by the homogeneous variables $(c_i, p_i)$ ($i=\theta, \sigma, \delta$), which describe again the connection and the densitized triad, and $(\phi, p_\phi)$, which describe the massless scalar field, and the inhomogeneous variables $(a^\xi_m,a^{\xi*}_m)$ and $(a^\varphi_m,a^{\varphi*}_m)$ (the latter if a matter scalar field has been introduced). In general relativity, they are subject to the Hamiltonian constraint
\begin{align}\label{eq:hamiltonianoGowdy}
\mathcal{C}_{GM} =& -\frac{1}{\kappa\gamma}(c_\theta p_\theta c_\sigma p_\sigma +c_\theta p_\theta c_\delta p_\delta+c_\sigma p_\sigma c_\delta p_\delta  ) \nonumber\\
& +\frac{(c_\sigma p_\sigma+c_\delta p_\delta)^2}{16\pi\gamma^2 p_\theta}H_{\mathrm{int}} + 2\pi p_\theta H_0 + \frac{p_\phi^2}{2}.
\end{align}
Here, $H_0=H_0^\xi+H_0^\varphi$, $H_{\mathrm{int}}=H_{\mathrm{int}}^\xi+H_{\mathrm{int}}^\varphi$, and
\begin{align}\label{eq:H0}
H^r_0 &= \sum\limits_{m=1}^\infty{m(a^{r*}_m a^r_m +a^{r*}_{-m} a^r_{-m} )}, \\
\label{eq:Hint}
H^r_{\mathrm{int}} &= \sum\limits_{m=1}^\infty{\frac{1}{m}(a^{r*}_m a^r_m +a^{r*}_{-m} a^r_{-m} + a^{r*}_m a^{r*}_{-m} +a^r_m a^r_{-m})},
\end{align}
where $r=\xi,\varphi$. The inhomogeneous degrees of freedom must also satisfy another constraint, that ensures that the total momentum of the system vanishes. Finally, it is easy to see that $0\leq H_{\mathrm{int}}\leq 2 H_0$ \cite{brizuela2011effective1,brizuela2011effective2}.

It is worth mentioning that the effective Hamiltonian constraint can be obtained from the original classical one by means of the substitution
\begin{equation}\label{eq:sustitucion}
c_i \rightarrow \frac{\sin{(\bar{\mu}_i c_i)}}{\bar{\mu}_i}, \ \mathrm{with} \   \bar{\mu}_i=\sqrt{\frac{\Delta p_i}{p_j p_k}}, \ i\neq j\neq k.
\end{equation}
The effective equations of motion can now be derived from the effective Hamiltonian by simply taking Poisson brackets with the elementary variables.

Using this effective Hamiltonian approach, Singh has recently studied the Bianchi I model in LQC to elucidate whether all the singularities are resolved~\cite{singh2012curvature}. The case with a massive scalar field as the matter content---in which inflation can occur---has also been thoroughly analyzed~\cite{gupt2013inflation}. The existence of anisotropies confers particular interest to the Bianchi I model, which exhibits a wide variety of singularities~\cite{maccallum1971singularities} that result in a rich behavior of the non-singular effective theory~\cite{gupt2012Kasner}. Furthermore, if one appeals to the Belinskii-Khalatnikov-Lifshitz conjecture \cite{BKL}, cosmological singularities would locally behave like homogeneous anisotropic regions of spacetime. Therefore, discussing whether LQC cures those singularities may lead to a general result about singularitfy resolution by loop quantum effects \cite{BKLloop}. In particular, it is interesting to analyze whether any persisting singularity is strong (implying the destruction of any free in-falling observable) or weak (which would let strong objects pass through). This strong or weak character is related with the extendibility of geodesics \cite{effectsingh}.

Since cosmological singularities are associated with divergences of the spacetime curvature, Singh studied these curvature invariants for the Bianchi I model with matter and a vanishing anisotropic stress \cite{singh2012curvature}. He found that, despite the fact that the directional Hubble rates, the expansion scalar, the shear scalar and the energy density are bounded in this model, the Ricci scalar $R$, the Kretschmann scalar $K = R_{\alpha\beta\mu\nu}R^{\alpha\beta\mu\nu}$ ($R_{\alpha\beta\mu\nu}$ being the Riemann tensor), and the square of the Weyl curvature $C_{\alpha\beta\mu\nu}C^{\alpha\beta\mu\nu}$ diverge if the volume vanishes and/or the pressure blows up at a finite energy density (as usual, spacetime indices are denoted with Greek letters).

We are interested in extending Singh's analysis to cosmologies with inhomogeneities. As we have already commented, we will focus on the three-torus Gowdy model with linearly polarized gravitational waves. By analyzing this relatively simple inhomogeneous model, we want to discuss the effect of inhomogeneities in the bouncing mechanism of LQC and thus check whether the results obtained in homogeneous LQC are still valid in their presence.
In particular, we want to explore the answer to the following questions: are the cosmological singularities still resolved by means of a Big Bounce when inhomogeneities come into the scene?; and, if this happens to be the case, how the inhomogeneities affect the bouncing process?

The structure of the rest of the work is as follows. Section \ref{sec:gowdy} summarizes the assumed effective Hamiltonian description and derives the corresponding equations of motion for the Gowdy model. In doing so, we extend the case analyzed in Refs.~\cite{brizuela2011effective1,brizuela2011effective2} by two means: first, by adopting an improved dynamics scheme of the form \eqref{mui}, and second by adding a matter content. The possible existence of singularities in our model is studied in Sec. \ref{sec:singularities}, determining the conditions under which the curvature invariants of the Bianchi I effective background may diverge. We find that the directional Hubble parameters, the expansion, the shear scalar, and the energy density are bounded, as in the purely homogeneous case, and that curvature invariants may diverge only if the physical (homogeneous) volume vanishes. Interestingly, we also show that the Hamiltonian constraint actually rules out the possibility that this volume becomes zero. Then, we concentrate on the possibility of a bounce in the Bianchi I volume in the process of avoidance of singularities. We determine in Sec. \ref{sec:bounce} the region of phase space in which this bounce may occur.  Moreover, we demonstrate that the value of the volume at this bounce is larger than the one corresponding to the purely homogeneous case, i.e., the inhomogeneities increase the volume at the bounce. Finally, Sec. \ref{sec:conclusions} concludes the work, summarizing the results and proposing some directions for future research.

\section{Effective dynamics of the Gowdy model}\label{sec:gowdy}

As we have pointed out, it is commonly assumed that the effective Hamiltonian constraint of the Gowdy model can be obtained from Eq.~\eqref{eq:hamiltonianoGowdy} with the replacement \eqref{eq:sustitucion}. The resulting expression can be written in a much more convenient form in terms of the variables
\begin{equation}\label{eq:variables_b_lambda_v}
b_{i}=\bar{\mu}_{i}c_{i}, \quad \lambda_{i}=\sqrt{16 \pi p_{i}}, \quad v=\sqrt{p_{\theta}p_{\sigma}p_{\delta}}.
\end{equation}
Note that, since $(16\pi)^{3/2}v=\lambda_{\theta}\lambda_{\sigma}\lambda_{\delta}$, not all these new variables are independent. Notice also that, up to a constant factor, $v$ provides the physical homogeneous volume of the three-torus sections (i.e., the physical volume if no inhomogeneities are present). The Poisson brackets of these variables can be easily deduced from Eq.~\eqref{eq:poissonbrackets}:
\begin{gather}\label{eq:newpoissonbrackets}
\{b_{i},\lambda_{j}\} = \frac{\Upsilon}{v}\lambda_{i}\delta_{ij}, \quad \{b_{i},v\}=\Upsilon, \nonumber\\ \{b_{i},b_{j}\}=\frac{\Upsilon}{v}(b_{i}-b_{j}),
\end{gather}
where $\Upsilon=\kappa\sqrt{\Delta }/2$. The Poisson brackets of other combinations vanish. The effective Hamiltonian constraint reads in these variables
\begin{eqnarray}\label{eq:hamiltonianoGowdy2}
\mathcal{C}^{\text{eff}}_{\text{GM}}&=&-\frac{v^{2}}{\kappa \gamma\Delta}\left[\sin b_{\theta}\sin b_{\sigma}+\sin b_{\sigma}\sin b_{\delta}+\sin b_{\delta}\sin b_{\theta}\right]  \nonumber \\
&& +\frac{v^{2}(\sin b_{\sigma}+\sin b_{\delta})^{2}}{\gamma^{2}\Delta\lambda^{2}_{\theta}}H_{\mathrm{int}}+\frac{\lambda_{\theta}^{2}}{8} H_{0}+\frac{p_{\phi}^{2}}{2}.
\end{eqnarray}

The evolution equations are obtained taking Poisson brackets with the effective Hamiltonian. Since it does not depend on $\phi$, it is immediate to see that $p_\phi=d\phi/d\tau$ is a constant of motion. Here, $\tau$ is an evolution parameter conjugate to the Hamiltonian. In fact, when no inhomogeneities are present it corresponds to a choice of lapse equal to the volume, up to numerical factors, namely $d\tau=\pi^2dt/(2v)$ in Eq.~\eqref{metricBI} \cite{garay2010inhomogeneous}. As for the other variables,
\begin{gather}
\label{ec:ec_mov_lambdatheta}
\frac{d \lambda_{\theta}}{d \tau} =  \frac{v\lambda_{\theta}}{2
\gamma\sqrt{\Delta}}\cos b_{\theta}(\sin b_{\sigma}+\sin b_{\delta}),
\end{gather}
\begin{widetext}
\begin{eqnarray}
\label{ec:ec_mov_lambdasigma}
\frac{d \lambda_{\sigma}}{d \tau} &=& \frac{v\lambda_{\sigma}}{2 \gamma\sqrt{\Delta}}\cos b_{\sigma}(\sin b_{\delta}+\sin b_{\theta})
- \frac{2\Upsilon v\lambda_{\sigma}}{ \gamma^{2}\Delta\lambda^{2}_{\theta}}\cos b_{\sigma}(\sin b_{\sigma}+\sin b_{\delta})H_{\mathrm{int}},
\\
\frac{d v}{d \tau} &=&
\frac{v^{2}}{2\gamma \sqrt{\Delta  }}\left[\sin b_{\theta}(\cos b_{\sigma}+\cos b_{\delta})+\sin b_{\sigma}(\cos b_{\delta}+\cos b_{\theta})+\sin b_{\delta}(\cos b_{\theta}+\cos b_{\sigma})\right]\nonumber\\
&&- \frac{2\Upsilon v^2}{ \gamma^{2}\Delta\lambda^{2}_{\theta}}(\sin b_{\sigma}+\sin b_{\delta})(\cos b_{\sigma}+\cos b_{\delta})H_{\mathrm{int}}, \\
\label{ec:ec_mov_btheta}
\frac{d b_{\theta}}{d \tau} &=& -\frac{v}{\gamma\sqrt{\Delta}}\bigg[\frac{1}{2}[(b_{\theta}-b_{\sigma})\cos b_{\sigma}(\sin b_{\theta}+\sin b_{\delta})+(b_{\theta}-b_{\delta})\cos b_{\delta}(\sin b_{\theta}+\sin b_{\sigma})]+\sin b_{\theta}\sin b_{\sigma}+\sin b_{\sigma}\sin b_{\delta} \nonumber\\
&& + \sin b_{\delta}\sin b_{\theta}\bigg]+\frac{2\Upsilon v}{ \gamma^{2}\Delta\lambda^{2}_{\theta}}(\sin b_{\sigma}+\sin b_{\delta})\left[(b_{\theta}-b_{\sigma})\cos b_{\sigma}+(b_{\theta}-b_{\delta})\cos b_{\delta}\right]H_{\mathrm{int}}+\frac{\Upsilon \lambda_{\theta}^{2}}{4 v}H_{0},\\
\label{ec:ec_mov_bsigma}
\frac{d b_{\sigma}}{d \tau} &=& -\frac{v}{\gamma\sqrt{\Delta}}\bigg[\frac{1}{2}[(b_{\sigma}-b_{\theta})\cos b_{\theta}(\sin b_{\sigma}+\sin b_{\delta})+(b_{\sigma}-b_{\delta})\cos b_{\delta}(\sin b_{\theta}+\sin b_{\sigma})]+\sin b_{\theta}\sin b_{\sigma}+\sin b_{\sigma}\sin b_{\delta} \nonumber\\
&& +\sin b_{\delta}\sin b_{\theta}\bigg] + \frac{2\Upsilon v}{ \gamma^{2}\Delta\lambda^{2}_{\theta}}(\sin b_{\sigma}+\sin b_{\delta})\left[\sin b_{\sigma}+\sin b_{\delta}+ (b_{\sigma}-b_{\delta})\cos b_{\delta}\right]H_{\mathrm{int}},
\\
\label{ec:ec_mov_am}
\frac{d a_{m}^{r}}{d \tau} &=&  -i \frac{v^{2}}{ \gamma^{2}\Delta\lambda^{2}_{\theta}|m|}(\sin b_{\sigma}+\sin b_{\delta})^{2}\big(a_{m}^{r}+a_{-m}^{r\ast}\big) -\frac{i}{8}\lambda_{\theta}^{2}|m|a_{m}^{r},
\end{eqnarray}
\end{widetext}
where the superindex $r$ denotes again both $\xi$ and $\varphi$. The dynamical equations for $\lambda_{\delta}$ and $b_{\delta}$ can be obtained from those of $\lambda_{\sigma}$ and $b_{\sigma}$ by interchanging $\sigma$ and $\delta$, and that for $a_m^{r\ast}$ is just the complex conjugate of Eq.~\eqref{ec:ec_mov_am}. With all these equations, it is straightforward to check that the quantity $C=v(b_{\sigma}-b_{\delta})$ is a constant of motion, as well as $x_{m}^{r}=a^{r}_{m}a^{r\ast}_{m}-a^{r}_{-m}a^{r\ast}_{-m}$ for every $m\in \mathbb{N}$. We can then simplify in part the dynamical equations by introducing the notation
\begin{equation}
b_{+}=\frac{b_{\sigma}+b_{\delta}}{2},  \qquad  b_{-}=\frac{b_{\sigma}-b_{\delta}}{2}, \qquad C=2v b_{-}.
\end{equation}
The effective Hamiltonian constraint adopts the form
\begin{equation}\label{eq:hamiltonianoGowdyfinal}
\begin{aligned}
\mathcal{C}_{\text{GM}}^{\text{eff}}=
&-\frac{v^{2}}{\kappa\gamma\Delta }
\left[2\sin b_{\theta} \sin b_{+} \cos \frac{C}{2v} + \cos^2{\frac{C}{2v}} - \cos^2{b_{+}} \right] \\
& +\frac{4 v^{2}}{\gamma^{2}\Delta \lambda_{\theta}^{2}}\sin^{2}b_{+}\cos^{2}\frac{C}{2v}H_{\mathrm{int}}
+\frac{\lambda_{\theta}^{2}}{8}H_{0}
+\frac{p^{2}_{\phi}}{2},
\end{aligned}
\end{equation}
and the homogeneous volume satisfies
\begin{eqnarray}\label{ec:ec_mov_v}
\frac{d v}{d \tau} &=&  \frac{v^{2}}{2 \gamma \sqrt{\Delta }}\left[2\sin(b_{+}+b_{\theta})\cos \frac{C}{2v}+ \sin 2b_{+}\right] \nonumber\\
&& - \frac{4\Upsilon v^2}{ \gamma^{2} \Delta \lambda^{2}_{\theta}}\sin 2b_{+}\cos^{2}\frac{C}{2v} H_{\mathrm{int}}.
\end{eqnarray}
The rest of the expressions can be found in the Appendix.

\section{Singularities in the Bianchi I and Gowdy models}\label{sec:singularities}

In this section, we study the possible singularities of the background Bianchi I metric, regarding the (gravitational and matter) inhomogeneities as the content of the spacetime~\cite{note1}. We assume that the energy contribution of the inhomogeneities remains finite at all times, so any divergence can arise only from the background.

In a spacetime singularity, curvature invariants typically diverge (although the divergence may arise in fact in some of their derivatives). In the framework of LQC, many of these curvature singularities are resolved in the homogeneous and isotropic case \cite{effectsingh}. As we have already mentioned, for the anisotropic Bianchi I model it was shown in Ref.~\cite{singh2012curvature} that, even if the Hubble parameters, the expansion, the shear scalar, and the energy density are bounded in effective LQC, the curvature invariants might still diverge under some conditions, a fact that implies that certain singularities might persist. In this section we will check whether such a divergent behavior is or not possible when inhomogeneities of the Gowdy type are present.

\subsection{Hubble parameters, expansion, shear scalar, and energy density}

It is easy to see from Eqs.~\eqref{eq:p_i} and \eqref{eq:variables_b_lambda_v} that the directional Hubble rate $H_\theta=\dot{a}_\theta/a_\theta$ is given by
\begin{equation}
H_\theta = \frac{\dot{\lambda}_\sigma}{\lambda_\sigma} + \frac{\dot{\lambda}_\delta}{\lambda_\delta} - \frac{\dot{\lambda}_\theta}{\lambda_\theta}.
\end{equation}
The expressions for $H_\sigma$ and $H_\delta$ can be obtained from this one by permutations of $(\theta,\delta,\sigma)$. Let us analyze the ratios $\dot{\lambda}_i/\lambda_i$ using the equations of motion of Sec. \ref{sec:gowdy}. For simplicity, we will work with a choice of time other than $\tau$, in order to absorb a factor of $v$ from all the dynamical equations. We choose the time $dT=v d\tau=\pi^2dt/2$ and denote in all our equations the derivative with respect to $T$ with an overdot.

Recalling that sines and cosines have always an absolute value not greater than the unit, it is straightforward to prove that the equation of motion of $\lambda_\theta$ \eqref{ec:ec_mov_lambdatheta} guarantees that $\dot{\lambda}_\theta/\lambda_\theta$ is bounded. This means that the derivative of $ \ln{\lambda_\theta}$ is bounded; so, if the evolution starts with $\lambda_\theta \neq 0$, the value $\lambda_\theta = 0$ cannot be reached in a finite amount of time $T$. In other words, $\lambda_\theta$ is bounded away from zero. For the same reason, it is also bounded away from infinity.

Now, employing the equations of motion \eqref{ec:ec_mov_lambdasigma} for $\lambda_\sigma$, one can express $\dot{\lambda}_\sigma/\lambda_\sigma$ as a bounded term plus the factor $H_{\mathrm{int}}/\lambda_\theta^2$ multiplied by another bounded term.
Assuming that $H_{\mathrm{int}}$ is finite, as it should happen in solutions with a small (and hence bounded) energy contribution of the inhomogeneities, and recalling that $\lambda_\theta$ cannot vanish in a finite time, we conclude that $\dot{\lambda}_\sigma/\lambda_\sigma$ cannot diverge in a finite time. Consequently, $\lambda_\sigma$ cannot vanish nor diverge to infinity in a finite period of time $T$ if the system begins with a nonvanishing initial value for this quantity. The same conclusion is reached for $\dot{\lambda}_\delta/\lambda_\delta$.

The behavior of the terms $\dot{\lambda}_i/\lambda_i$ guarantees that the directional Hubble parameters of the Bianchi I part of the Gowdy model are locally bounded on each effective trajectory, similarly to what occurs in the genuine Bianchi I case, without gravitational and matter inhomogeneities.

Let us now analyze the behavior of the expansion and the shear scalar. The expansion scalar $\bar{\theta}=\dot{v}/v$ can be easily calculated using Eqs.~\eqref{eq:variables_b_lambda_v} and \eqref{eq:p_i}. One gets
\begin{equation}
\bar{\theta}=\frac{\dot{v}}{v}=H_\theta+H_\sigma+H_\delta.
\end{equation}
Therefore, clearly, $\bar{\theta}$ is always bounded on effective trajectories. Besides, $v$ cannot vanish nor diverge to infinity in a finite time $T$. In fact, not even can $v$ vanish asymptotically, as we will see below.

On the other hand, the expansion tensor can be expressed as \begin{equation}\label{eq:expansiontensor}
\bar{\theta}_{\mu\nu}=\sigma_{\mu\nu}+\frac{1}{3}\bar{\theta}\left(g_{\mu\nu}+v_\mu v_\nu\right),
\end{equation}
where $\sigma_{\mu\nu}$ is the (traceless) shear tensor, $g_{\mu\nu}$ is the metric tensor and $v^\mu$ is the unit velocity tangent to the time-like geodesics. Its covariant derivative is
$v_{\mu;\nu}=\bar{\theta}_{\mu\nu}$, since there is no vorticity in the system \cite{kramers}. Using this relation and Eq.~\eqref{eq:expansiontensor}, the shear scalar $\sigma^2=\sigma_{\mu\nu}\sigma^{\mu\nu}/2$ can be calculated, obtaining:
\begin{equation}
\sigma^2=\frac{1}{6}\left[\left(H_\theta-H_\sigma\right)^2 + \left(H_\sigma-H_\delta\right)^2 + \left(H_\delta-H_\theta\right)^2\right].
\end{equation}
Again, we immediately see that this scalar is locally bounded on each effective solution, because the same happens with all the Hubble parameters.

Finally, the energy density is defined as $\rho= \mathcal{C}_{{\rm mat}}/v^2$, where $\mathcal{C}_{{\rm mat}}$ is the part of the Hamiltonian that includes the inhomogeneities and the scalar field contribution (regarded all as matter content). It can be calculated from the Hamiltonian constraint $ \mathcal{C}_{\text{GM}}^{\text{eff}}=0$, with the appropriate densitization. Using Eq.~\eqref{eq:hamiltonianoGowdyfinal} divided by $v^2$, one gets
\begin{equation}
\rho=\frac{1}{\kappa\gamma\Delta }
\left[2\sin b_{\theta} \sin b_{+} \cos \frac{C}{2v} + \cos^2{\frac{C}{2v}} - \cos^2{b_{+}} \right],
\end{equation}
which is also a bounded quantity.

In conclusion, the presence of inhomogeneities, viewed as a matter content in the Bianchi I background part of the system, does not alter the fact that the Hubble parameters, the expansion, the shear scalar, and the energy density remain bounded in the effective dynamical evolution, provided that $H_{\mathrm{int}}$ is kept finite.

\subsection{Curvature invariants}

We now analyze the behavior of the Ricci scalar $R$, the Kretschmann scalar $K$, and the square of the Weyl curvature, in order to discuss whether it is possible that they blow out in the evolution. These scalars can be written in terms of the Hubble rates and the second derivatives of the directional scale factors \cite{singh2012curvature}. We have
\begin{align}\label{eq:Ricci}
R &= 2\left[H_\theta H_\sigma+H_\sigma H_\delta+H_\delta H_\theta+\sum\limits_{i}^{}\frac{\ddot{a}_i}{a_i}\right],
\\
\label{eq:Kretschmann}
K &= 4\left[H_\theta^2 H_\sigma^2+H_\sigma^2 H_\delta^2+H_\delta^2 H_\theta^2+\sum\limits_{i}^{}\frac{\ddot{a}_i^2}{a_i^2}\right],
\end{align}
whereas the square of the Weyl curvature is
\begin{align}
\label{eq:Weyl}
& C_{\alpha\beta\mu\nu}C^{\alpha\beta\mu\nu} = \frac{4}{3}\frac{\ddot{a}_\theta}{a_\theta}\left(\frac{\ddot{a}_\theta}{a_\theta}-H_\theta H_\sigma+2H_\sigma H_\delta-H_\delta H_\theta\right) \nonumber \\
&-\frac{4}{3}\bigg[\frac{\ddot{a}_\theta}{a_\theta}-\frac{1}{2}H_\theta^2 (H_\sigma-H_\delta)^2\bigg]+(\text{cyclic permutations}).
\end{align}
As indicated, the rest of the terms are obtained from the explicit ones with cyclic permutations of $(\theta,\sigma,\delta)$.
For the studied Gowdy model, we have seen that the Hubble parameters are locally bounded on each trajectory, so the question of whether these curvature invariants diverge depends exclusively on the behavior of the terms $\ddot{a}_i/a_i$. The form of these terms can be obtained after some algebraic manipulations using Eqs.~\eqref{eq:p_i} and \eqref{eq:variables_b_lambda_v} \cite{singh2012curvature}. Thus,
\begin{equation}
\frac{\ddot{a}_\theta}{a_\theta} = \frac{\ddot{\lambda}_\sigma}{\lambda_\sigma}+\frac{\ddot{\lambda}_\delta}{\lambda_\delta}-
\frac{\ddot{\lambda}_\theta}{\lambda_\theta}+\frac{1}{2}\left(H_\theta^2-H_\theta H_\sigma+H_\sigma H_\delta-H_\delta H_\theta\right).
\end{equation}
The expressions for $\ddot{a}_\sigma/a_\sigma$ and $\ddot{a}_\delta/a_\delta$ can be obtained by permutations of the labels $(\theta,\sigma,\delta)$.

Hence, the boundedness of the curvature invariants turns out to be determined by that of the terms $\ddot{\lambda}_i/\lambda_i$. The behavior of these terms can be deduced from the equations of motion \eqref{ec:ec_mov_lambdatheta} and \eqref{ec:ec_mov_lambdasigma} of the variables $\lambda_i$. The resulting expressions are
\begin{widetext}
\begin{eqnarray}
\frac{\ddot{\lambda}_\theta}{\lambda_\theta}&=&\frac{1}{2\gamma\sqrt{\Delta}}\left[-\dot{b}_\theta\sin{b_\theta}(\sin{b_\sigma}
+\sin{b_\delta})
+\cos{b_\theta}(\dot{b}_\sigma\cos{b_\sigma}+\dot{b}_\delta\cos{b_\delta})\right]+\frac{1}{4\gamma^2\Delta}\cos^2{b_{\theta}}(\sin b_{\sigma}+\sin b_{\delta})^2,\\
\frac{\ddot{\lambda}_\sigma}{\lambda_\sigma} &=& \frac{1}{2\gamma\sqrt{\Delta}}\left[-\dot{b}_\sigma\sin{b_\sigma}(\sin{b_\delta}+\sin{b_\theta})+\cos{b_\sigma}
(\dot{b}_\delta\cos{b_\delta}+\dot{b}_\theta\cos{b_\theta})\right] - \frac{2\Upsilon}{\gamma^2\Delta\lambda_\theta^2}\cos{b_\sigma}(\sin{b_\sigma}
+\sin{b_\delta})\dot{H}_{\mathrm{int}} \nonumber\\
&&+\frac{2\Upsilon}{\gamma^2\Delta\lambda_\theta^2}\left\{2\frac{\dot{\lambda}_\theta}{\lambda_\theta}\cos{b_\sigma}(\sin{b_\sigma}+\sin{b_\delta}) + \left[\dot{b}_\sigma\sin{b_\sigma}
(\sin{b_\sigma}+\sin{b_\delta})-\cos{b_\sigma}(\dot{b}_\sigma\cos{b_\sigma}+\dot{b}_\delta\cos{b_\delta})\right]\right\}H_{\mathrm{int}}
\nonumber\\
&& +\cos^2 b_{\sigma}\left[\frac{1}{2\gamma\sqrt{\Delta}}(\sin b_\delta+\sin b_\theta)-\frac{2\Upsilon}{\gamma^2\Delta\lambda_\theta^2}(\sin b_\sigma+\sin b_\delta)H_{\mathrm{int}}\right]^2.
\end{eqnarray}
\end{widetext}
The expression for $\ddot{\lambda}_\delta/\lambda_\delta$ is similar and can obtained interchanging $\sigma$ and $\delta$ in the last equation. We have already proven that $\lambda_\theta$ cannot vanish at a finite time, nor $\dot{\lambda}_\theta/\lambda_\theta$ can diverge. Then, accepting that $H_0$ remains finite, so does $H_{\mathrm{int}} \leq 2H_0$, and also its derivative, which, after some manipulation of Eqs.~\eqref{eq:Hint} and \eqref{ec:ec_mov_am}, can be seen to satisfy the bound
\begin{equation}
 |\dot{H}_{\mathrm{int}}| \leq  \left(\frac{4v}{\gamma^2\Delta\lambda_\theta^2}(\sin b_\sigma+\sin b_\delta)^2+\frac{\lambda_\theta^2}{4v}\right)H_0.
\end{equation}
Incidentally, $|\dot{H}_{0}|$ has a similar bound, consistent with the assumption that $H_0$ remains finite for finite $T$. Thus, the only terms that might cause the divergence of curvature invariants are the derivatives of the variables $b_i$: $\dot{b}_\theta$, $\dot{b}_\sigma$, and $\dot{b}_\delta$. Their values are given by the equations of motion \eqref{ec:ec_mov_btheta}, \eqref{ec:ec_mov_bsigma}, and the corresponding equation for $\dot{b}_\delta$. Recalling again that $\lambda_\theta$ and $v$ cannot vanish nor diverge at a finite time, and provided that the energy of the inhomogeneities does not blow up, we conclude that the curvature invariants may diverge only if any of the variables $b_i$ diverges.

However, in fact none of the $b_i$'s can diverge. Let us suppose that a certain dynamical trajectory encounters a singularity at a certain time $T_{\mathrm{S}}$. In that trajectory, the functions $\lambda_i$ and $H_{\mathrm{int}}$ can be extended continuously beyond $T_{\mathrm{S}}$, owing to the boundedness of their derivatives up to that point. Consequently, one can safely admit that the derivatives $\dot{b}_i$, regarded as functions of $T$ and $\vec{b} =(b_\theta,b_\sigma,b_\delta)$, are continuous. Besides, one can check that they satisfy locally the Lipschitz condition in $\vec{b}$. Then, Picard's Existence Theorem \cite{picard} ensures that the $b_i$'s are bounded in a suitable interval of time that can be extended beyond $T_{\mathrm{S}}$. Therefore, there can not be a singularity at $T_{\mathrm{S}}$.

In more detail, we can start with general initial conditions $(T_0,\vec{b}_0) \in \mathbb{R}^4$ and two real numbers $L$ and $b$, with the only restriction that $|T_{\mathrm{S}}-T_0|<L$ and that, in the set
\begin{equation}
  B = \{ (T, \vec{b}) : |T - T_0| < L, \| \vec{b} - \vec{b}_0 \| < b \},
\end{equation}
the derivatives $\dot{b}_i$ satisfy the Lipschitz condition in $\vec{b}$. Actually, given our evolution equations, we can take $b$ as large as we want. Hence, Picard's theorem guarantees the existence of a solution $\vec{b}(T)$ in the interval $|T - T_0| < \min(L,b/M)$, where $M = \sup_B\|d\vec{b}(T,\vec{b})/dT\|$, satisfying the initial condition $\vec{b}(T_0)=\vec{b}_0$.
In our case, it is possible to see that $M$ depends in fact on $b$ and that the ratio $b/M$ has a non-vanishing limit for infinite large $b$, independent of the choice of $L$. Selecting now $L$ properly, so that it is smaller than the value of this limit (and $T_0$ accordingly, so that $|T_{\mathrm{S}}-T_0|<L$), we conclude that the regularity of the solution in a neighborhood of $T_{\mathrm{S}}$ is in fact secured.

Finally, let us show that in fact the homogeneous volume can never vanish, not even asymptotically, and so is bounded from below on each effective trajectory. If we set $v=0$ in the effective Hamiltonian constraint \eqref{eq:hamiltonianoGowdy2}, we obtain that $\mathcal{C}^{\text{eff}}_{\text{G}}=\lambda_{\theta}^{2}H_{0}/8+p_{\phi}^2/{2}$. Since the two contributions on the right-hand side of this expression are nonnegative, (as soon as we have a nonzero momentum $p_\phi$ of the homogeneous scalar field) the Hamiltonian constraint cannot be satisfied for vanishing volume. Therefore, in these circumstances the volume cannot vanish in the evolution, but instead remains bounded away from zero. As a consequence, no cosmological singularity with vanishing volume may exist in fact. This is a noticeable result, that generalizes to the inhomogeneous Gowdy model the resolution of the Big Bang singularity found in the homogeneous case.

\section{Feasibility of the bounce}\label{sec:bounce}

Supposing that there is a bounce in the homogeneous volume, we now delimitate the regions of phase space where it can occur. This event would be characterized by the vanishing of the derivative $\dot{v}$. Imposing this in Eq.~\eqref{ec:ec_mov_v}, taking into account that the volume is strictly positive, and provided that $\sin 2 b_+$ does not vanish (the case of vanishing sine can be analyzed as a limit), we obtain the following quadratic equation in $x=\cos(C/2v)$:
\begin{equation}\label{ec:ecuacion_x_simplif}
  x^{2}  -2 \beta S x  - \beta  = 0 ,
\end{equation}
where
\begin{equation}\label{eq:betaS}
\beta = \frac{\gamma}{4\kappa } \frac{\lambda^{2}_{\theta}}{H_{\mathrm{int}}} , \quad S = \frac{\sin(b_{+}+b_{\theta})}{\sin 2b_{+}}.
\end{equation}
The solutions of this equation are
\begin{equation}\label{eq:sol_x}
  x = \beta S \pm \sqrt{\beta^2S^2+\beta}.
\end{equation}
Note that the argument of the square root is always positive and hence the two solutions are always real. Moreover, since $\beta$ is positive, one of the solutions is positive, and the other one is negative. From now on, we will call p-branch and m-branch these solutions, respectively, corresponding to the plus and the minus sign in front of the square root.

\begin{figure*}
 \begin{center}
  \subfigure[]{\includegraphics[width=.329\linewidth]{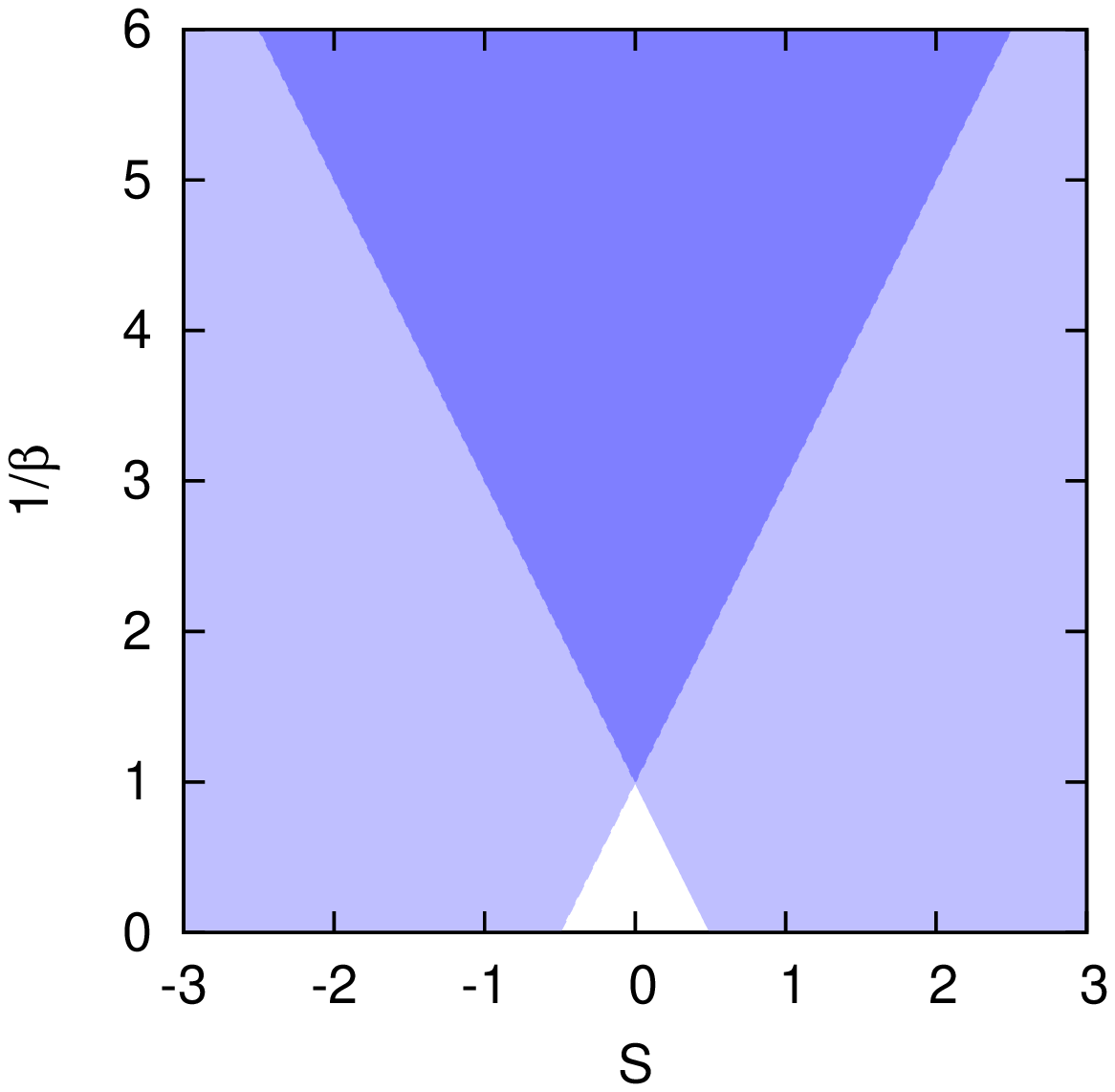}\label{fig:regionfactible1}}
  \subfigure[]{\includegraphics[width=.329\linewidth]{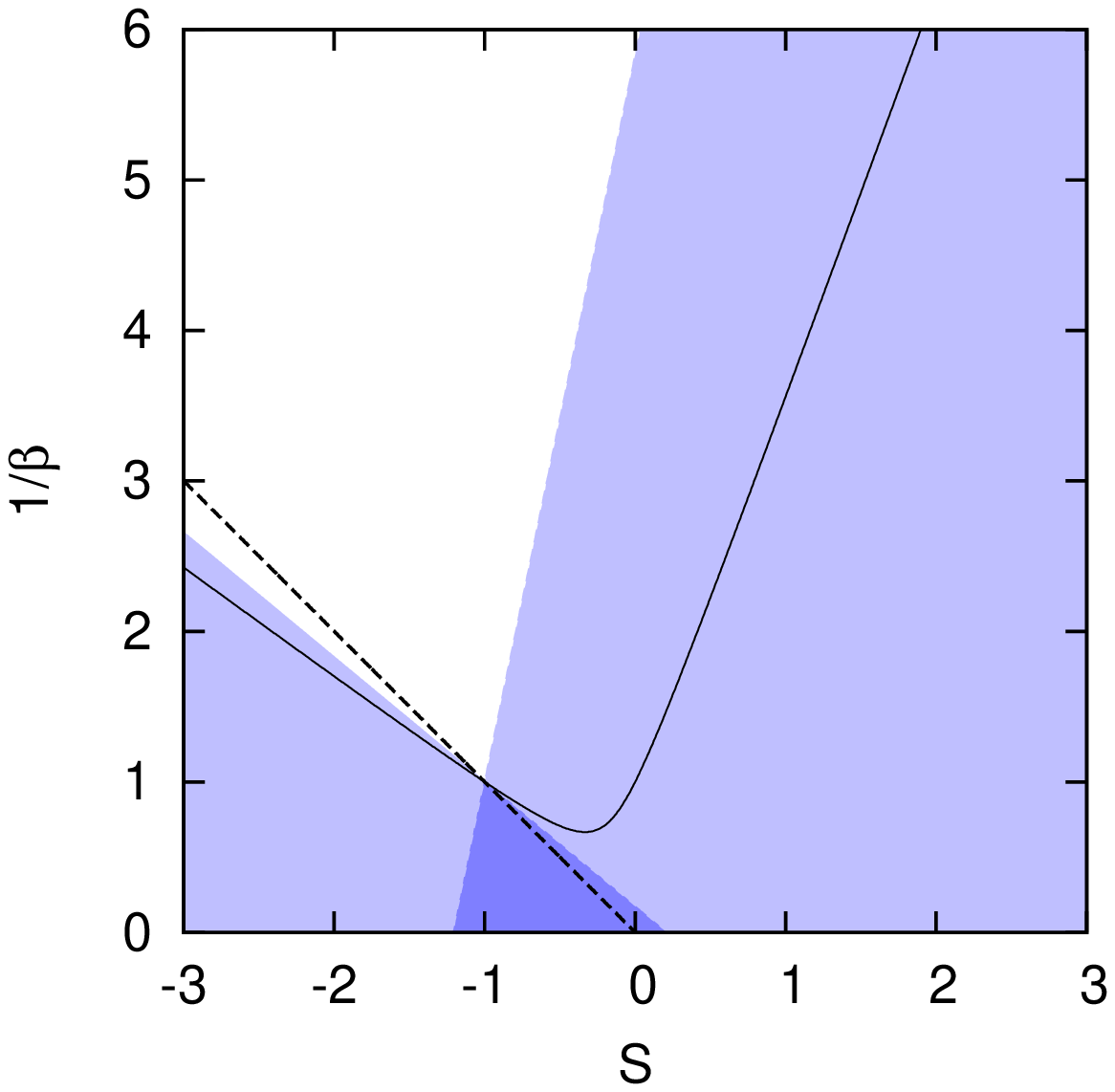}\label{fig:regionfactible2}}
  \subfigure[]{\includegraphics[width=.329\linewidth]{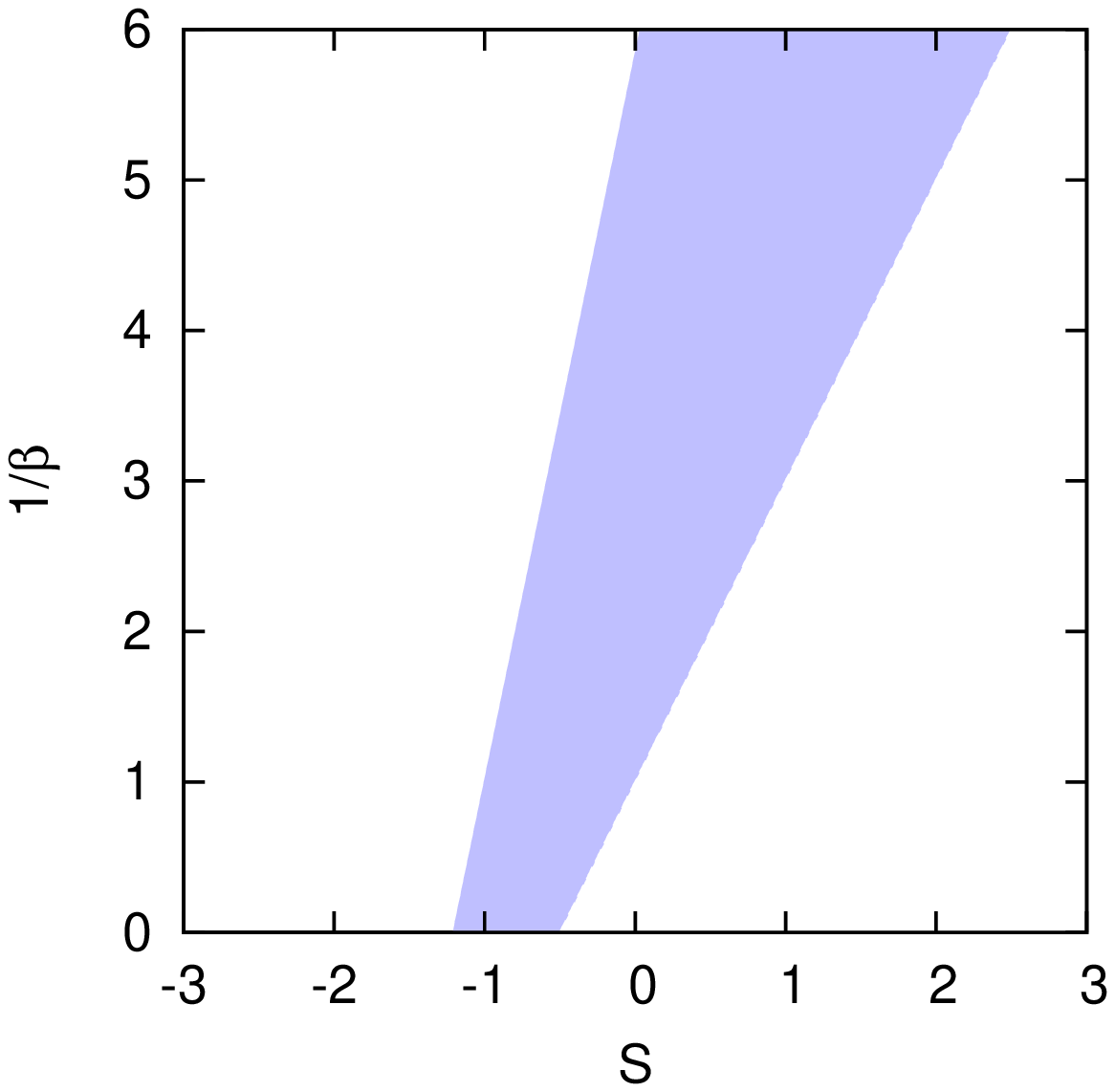}\label{fig:regionfactible1y2}}
  \caption{(a): The colored area is the region of the parameter space where there is at least one solution to Eq.~\eqref{ec:ecuacion_x_simplif} such that $|x| \leq 1$. The left (right) wedge corresponds to the p-branch (m-branch). Both branches coexist in the darker area. (b): Region compatible with the condition $v^2 > 0$ for $z=1$. The left (right) wedge corresponds to the m-branch (p-branch). The lower wedge admits both branches. The line $\beta S=-z$ and the hyperbola $\beta-1+2\beta S(\beta S+z)=0$ are also depicted. (c): Region where both restrictions are satisfied simultaneously.}\label{fig:regionfactible}
 \end{center}
\end{figure*}

Since $x$ is a cosine, only solutions satisfying $|x| \leq 1$ are valid. This implies the restriction
\begin{equation}
  1 \mp \beta S \geq \sqrt{\beta^2 S^2 + \beta},
\end{equation}
which leads to the equivalent one
\begin{equation}\label{ec:regionfactible1}
  \frac{1}{\beta} \geq 1 \pm 2 S
\end{equation}
if $\beta \neq 0$. Figure~\ref{fig:regionfactible1} shows the regions of the $(S,1/\beta)$ plane where these inequalities are satisfied.

The special limits $\beta \rightarrow \infty$, which corresponds to homogeneity (vanishing $H_{\mathrm{int}}$), and $S \rightarrow \pm\infty$ can be easily analyzed separately. In each case, there is only one solution, namely $x=-1/(2S)$ if $\beta \rightarrow \infty$ and $x=0$ if $S \rightarrow \pm\infty$ [assuming $\sin(b_+ + b_\theta) \neq 0$].

There are additional restrictions on the values of $\beta$ and $S$ arising from the imposition of the Hamiltonian constraint, $\mathcal{C}_{\text{GM}}^{\text{eff}}=0$. Using Eq.~\eqref{eq:hamiltonianoGowdyfinal} and the solutions~\eqref{eq:sol_x}, we obtain a linear equation in $v^2$, whose solution is
\begin{equation}\label{ec:sol_v2}
v^2 =\frac{\kappa\gamma P\Delta}{\beta - 1 + 2(\beta S + z) (\beta S \pm \sqrt{\beta^2 S^2+\beta})},
\end{equation}
where
\begin{equation}
P = \frac{1}{8}\lambda_\theta^2H_0+\frac{1}{2}p_\phi^2, \quad z= \frac{1}{2} \sin(b_{\theta} - b_{+})\tan b_{+}.
\end{equation}
For consistency, we must impose the positivity of the solutions, so that $v^2>0$. This constrains further the feasible region of our parameter space. The numerator on the right-hand side of Eq.~\eqref{ec:sol_v2} is clearly positive (recall, in particular, that so is $H_0$). Therefore, the denominator must be positive as well, i.e., we have to require
\begin{equation}\label{eq:regionfactible2}
   \beta - 1 + 2 \beta S (\beta S + z) \geq \mp 2 (\beta S + z) \sqrt{\beta^2 S^2 + \beta}.
\end{equation}
If the inequality is saturated, the Hamiltonian constraint can not be satisfied, since $\mathcal{C}_{\text{GM}}^{\text{eff}}=P>0$, irrespective of the value of $v^2$. This happens when either of the following equalities hold:
\begin{align}
  \frac{1}{\beta} &= 1 + 2 \left(z + \sqrt{z^2 + 1}\right) (S + z), \\
  \frac{1}{\beta} &= 1 + 2 \left(z - \sqrt{z^2 + 1}\right) (S + z).
\end{align}
These lines split the half-plane $(S,1/\beta)$ in four wedges [see Fig.~\ref{fig:regionfactible2}]. In the upper and the lower wedges
\begin{equation}\label{eq:uplowwedge}
  |\beta - 1 + 2 \beta S (\beta S + z)| > \left|2 (\beta S + z)\sqrt{\beta^2 S^2+\beta}\right|;
\end{equation}
so  Eq.~\eqref{eq:regionfactible2} holds for both signs in that inequality if, in addition, $\beta - 1 + 2 \beta S (\beta S + z)$ is positive. It can be seen that this condition removes the upper wedge. In conclusion, the p-branch and the m-branch lead to admissible solutions in the lower wedge, but none of them is acceptable in the upper one.

In the left and the right wedge, the inequality \eqref{eq:uplowwedge} is reversed. Therefore, the restriction \eqref{eq:regionfactible2} is now satisfied only for one sign on the right-hand side of that equation, depending on whether
\begin{equation}
  \mp (\beta S + z) < 0.
\end{equation}
This condition is satisfied for the m-branch in the left wedge and for the p-branch in the right one.

All together, conditions \eqref{ec:regionfactible1} and \eqref{eq:regionfactible2} are simultaneously met in the region characterized by
\begin{align}
 \frac{1}{\beta} &< 1 + 2 \left(z + \sgn(z)\sqrt{z^2+1}\right)(S+z),\label{eq:upperbound} \\
 \frac{1}{\beta} &\geq 1 + 2 \sgn(z) S, \\
 \frac{1}{\beta} &\neq 1 + 2 \left(z - \sgn(z)\sqrt{z^2+1}\right)(S+z).
\end{align}
An example of this kind of feasible region is shown in Fig.~\ref{fig:regionfactible1y2}. Besides, the conditions select a unique value of $x$ (belonging to the p-branch if $z>0$ or to the m-branch if $z<0$) at each point of the region.

\begin{figure*}
 \begin{center}
   \subfigure[]{\includegraphics[width=0.4\linewidth]{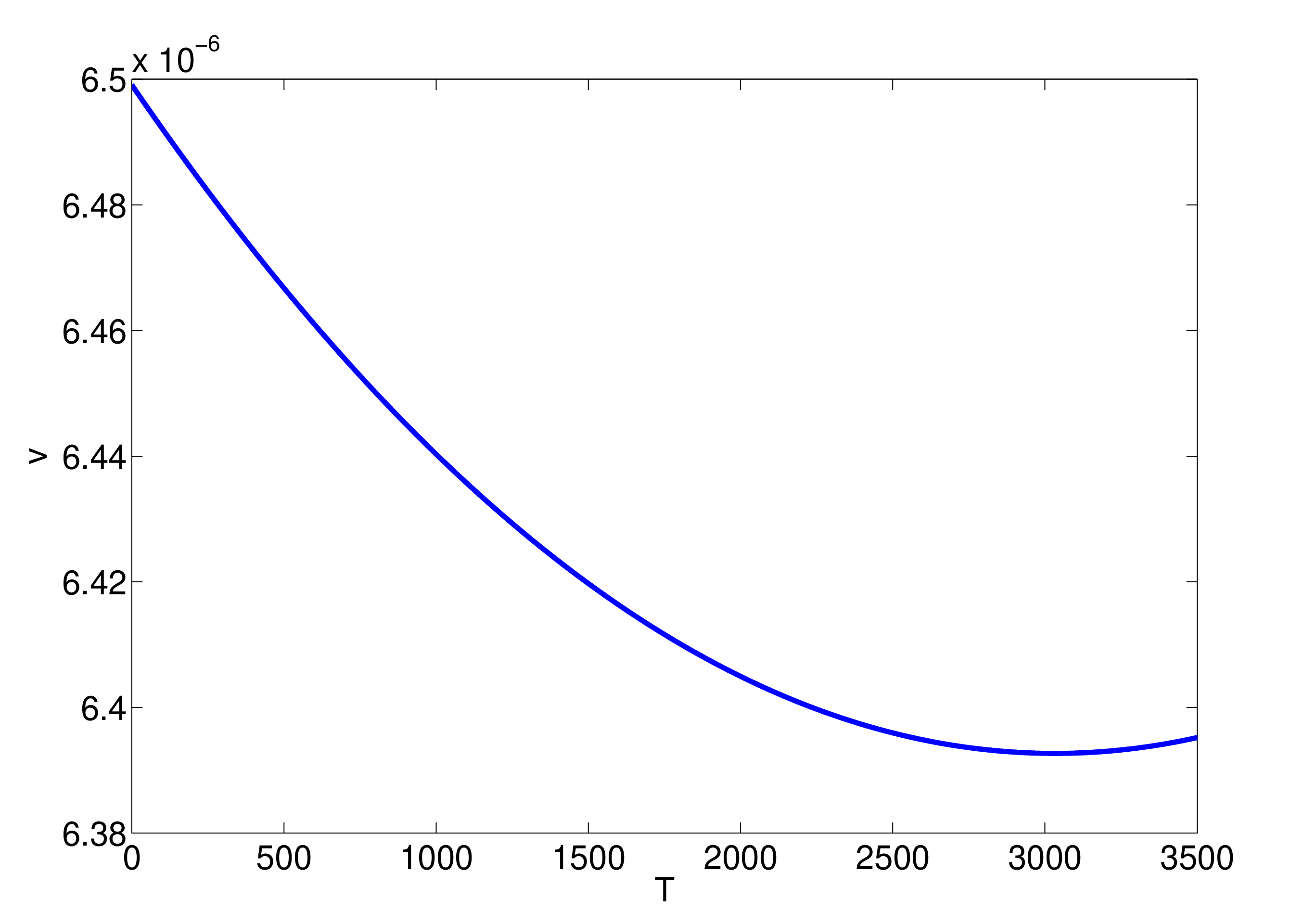}\label{fig:bounce1}}
   \subfigure[]{\includegraphics[width=0.4\linewidth]{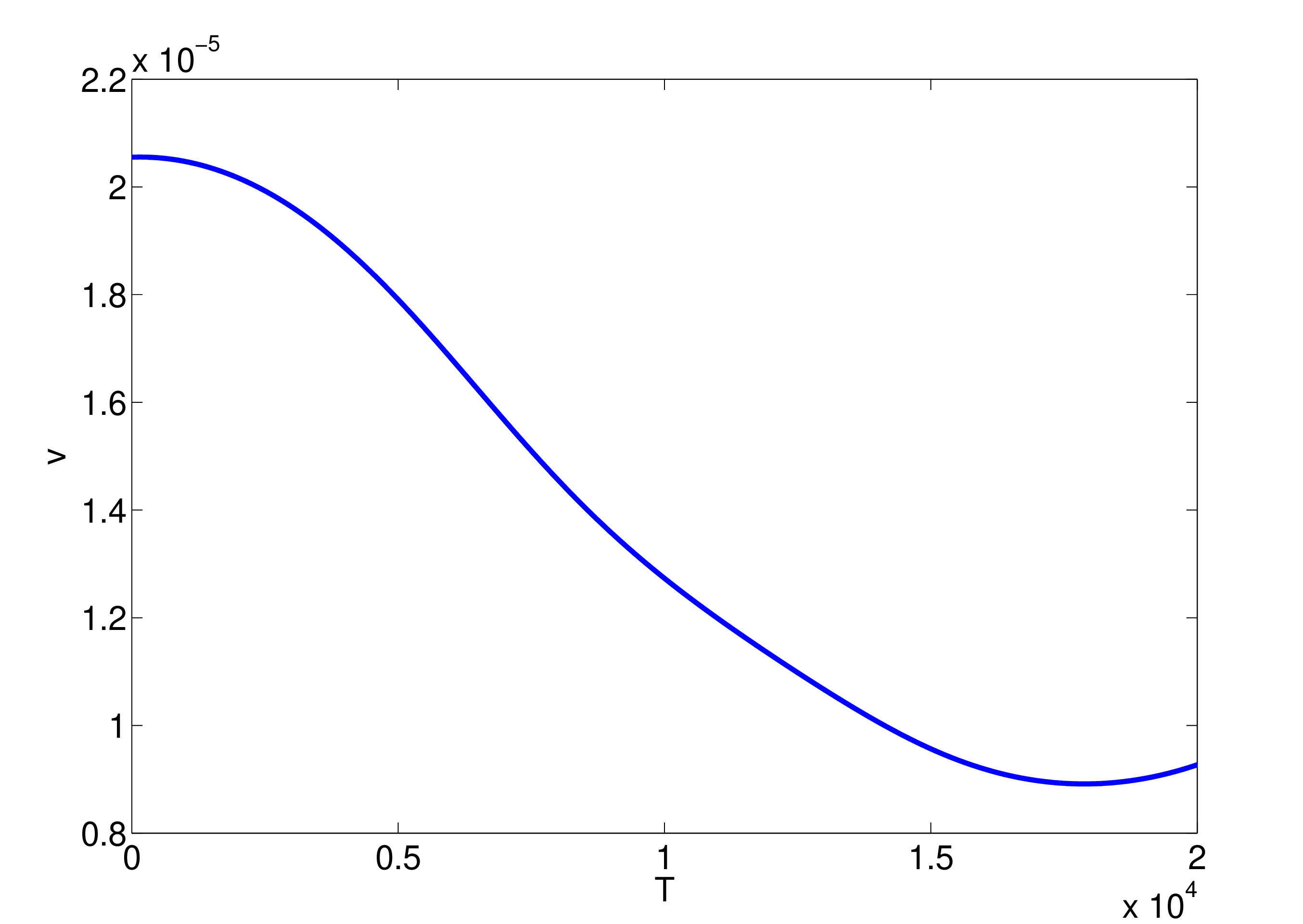}\label{fig:bounce2}}
  \caption{Evolution of the homogeneous volume $v$ in effective trajectories featuring bounces. The dynamical equations~\eqref{ec:ec_mov_lambdatheta_2}--\eqref{ec:ec_mov_am_2} were integrated numerically with Matlab, choosing the initial values so that the trajectory started in the feasible region: (a) $\beta=100$, $S=-0.994771$, $z=0.977429$; (b) $\beta=1000$, $S=-0.525495$, $z=0.0989956$. For simplicity, the only present inhomogeneous modes were those with $m=1,-1$ in both cases.}\label{fig:bounce}
 \end{center}
\end{figure*}

Therefore, there are large regions of the parameter space where a bounce in the homogeneous volume cannot occur. On the other hand, we have not proven that effective trajectories within the feasible region do undergo a bounce. Nevertheless, we have found that this can indeed be the case by integrating numerically the dynamical equations~\eqref{ec:ec_mov_lambdatheta_2}--\eqref{ec:ec_mov_am_2} with  Matlab in some simple situations with only a few inhomogeneous modes present. Figure~\ref{fig:bounce} displays two examples of bouncing trajectories. The reliability of the numerics was tested by changing the range of the integration step and tracking the Hamiltonian constraint, which keeps well below $10^{-12}$ at every point of the trajectories. 

We can use Eq.~\eqref{ec:sol_v2} to analyze how the inhomogeneities affect the value of the homogeneous volume at its bounce. The numerator on the right-hand side of the equation is clearly larger in the inhomogeneous ($H_0>0$) than in the homogeneous case ($H_0=0$). On the other hand, the denominator, which can be rewritten as $y(x) = x^2 + 2 z x - 1$ , reaches its minimum at $x=-z$. To be specific, let us consider the case $z > 0$, so the allowed values of $x$ belong to the p-branch (the analysis is essentially the same if $z < 0$). From Eq.~\eqref{eq:sol_x},
\begin{equation}
x = \frac{1}{\sqrt{S^2+\frac{1}{\beta}}-S}.
\end{equation}
One can easily check that, in this branch, $x$ decreases as $\beta$ diminishes---that is, as the inhomogeneities become more important. Since $y$ is monotonically increasing in the positive semi-axis (and $x>0$ in the p-branch), we conclude that the denominator is smaller when the inhomogeneities are stronger.

In conclusion, the inhomogeneities result in an increase of the numerator on the right-hand side of Eq.~\eqref{ec:sol_v2} and a decrease of the denominator. Therefore, the value of the homogeneous volume at the bounce is larger than the one corresponding to the homogeneous case. In other words, the inhomogeneities increase the value of this volume at the (potential) bounce. This can be understood heuristically by recalling that the contribution of the perturbations to the energy density is positive, according to Eq.~\eqref{eq:hamiltonianoGowdyfinal}. Consequently, it seems reasonable that the bouncing regime is entered at bigger volumes.

\section{Conclusions}\label{sec:conclusions}
In this work we have studied the possible existence of singularities in inhomogeneous and anisotropic cosmologies, focusing our attention on the three-torus Gowdy model, with linearly polarized gravitational waves and a minimally coupled scalar field with the same symmetries as the geometry. This model can be interpreted as an inhomogeneous content, made of the inhomogeneous part of the gravitational waves and the matter scalar field, which propagates in a homogeneous Bianchi I background. The physics of this simple model facilitates the comprehension of the role of inhomogeneities in quantum gravity and allows one to check whether the results obtained in homogeneous LQC are still valid in the presence of these inhomogeneities.

On the one hand, we have generalized the results derived by Singh \cite{singh2012curvature} for the Bianchi I model by studying the possible divergences of the curvature invariants in the considered inhomogeneous model. We have confirmed that, as in the Bianchi I case, the directional Hubble rates, the expansion, the shear scalar, and the energy density in the Bianchi background cannot diverge to infinity in the effective dynamics, provided that the energy interaction of the inhomogeneities, $H_\text{int}$, is kept finite. Moreover, in contrast to the homogeneous Bianchi I model (where certain pathologies are not completely discarded generally), we have proven that in this inhomogeneous model all the potential cosmological singularities are avoided, as long as the energy of the inhomogeneities, $H_0$ (recall that $H_0\geq H_\text{int}/2$), is finite. In particular, we have demonstrated that, thanks to the restrictions imposed by the Hamiltonian constraint, the homogeneous volume $v$ can never vanish in the evolution, and hence singularities with vanishing homogeneous volume cannot exist. This generalizes the results about the resolution of the Big Bang singularity found in homogeneous cosmologies to the considered inhomogeneous Gowdy model.

On the other hand, we have considered the possibility of a bounce in the homogeneous volume and determined the regions of the phase space where it may happen. We have found that the value of the volume at the bounce is given by the two possible branches of solutions to a quadratic equation. These two branches, however, are not generally valid in the complete phase space as a consequence of two consistency restrictions. The first is due to the fact that the equation of motion for the volume depends on the quantity $\cos(C/2v)$, whose absolute value must be less than or equal to one. The second restriction results from the positiveness of the square of the volume. Taking this into account, we have determined the feasible region for each of the branches and we have found that at most one of the branches may be valid (and this specific branch is not always the same). After characterizing the region of the phase space where the bounce in the homogeneous variable is feasible, we have studied numerically some effective trajectories within that region which do indeed undergo a bounce. Nonetheless, we have limited our study to simple situations with just a few inhomogeneous modes. Finally, we have analyzed the value of the feasible solutions to demonstrate that, provided a bounce in the homogeneous volume does occur, the inhomogeneities make the value of this volume at the bounce increase with respect to its counterpart in the homogeneous case.

Although a truncation of the model to a finite number of degrees of freedom is necessary to deal with it numerically, it would be interesting to extend the analysis of the effective trajectories presented in this work to scenarios with an increasing number of modes present. It is worth emphasizing that we have discussed the resolution of singularities of the Bianchi I (effective) background in the presence of the inhomogeneities corresponding to the gravitational waves and scalar field content of a Gowdy cosmology, but we have not fully investigated the possibility of singularities in the inhomogeneous part of the system. That investigation has to face as well the difficulties of dealing with an infinite number of degrees of freedom. Another topic for further discussion is the possible existence of bounces in other quantities, for instance in the anisotropies $\lambda_i$. As a more ambitious goal, numerical simulations could be used to confirm whether the effective dynamics is actually correct by studying the quantum evolution of appropriate physical states that, in the large volume regions, posses a semiclassical behavior with a small content of inhomogeneities. Again, a truncation of the model may be necessary, passing finally to the limit of an infinite number of modes.

\section*{Acknowledgements}

We would like to thank L. J. Garay, D. Mart\'{\i}n-de Blas, and J. Olmedo for discussions. This work was supported by the Projects No.\ MICINN/MINECO FIS2011-30145-C03-02 and CPAN CSD2007-00042. M. F.-M. acknowledges CSIC and the European Social Fund for support under the grant JAEPre\_2010\_01544. \linebreak

\appendix*

\section{Effective equations of motion}
In terms of the variables introduced at the end of Sec.~\ref{sec:gowdy},  $b_{+}=(b_{\sigma}+b_{\delta})/2$ and $b_{-}=C/(2v)$, the equations of motion read
\begin{equation}
\frac{d \lambda_{\theta}}{d \tau} = \frac{v\lambda_{\theta}}{\gamma\sqrt{\Delta}}\cos b_{\theta}\sin b_{+} \cos \frac{C}{2v},
\end{equation}
\begin{widetext}
\allowdisplaybreaks
\begin{eqnarray}\label{ec:ec_mov_lambdatheta_2}
\frac{d \lambda_{\sigma}}{d \tau} & = & \frac{v\lambda_{\sigma}}{2 \gamma\sqrt{\Delta}}\left[\sin b_{\theta} + \sin \left(b_{+}-\frac{C}{2v}\right)\right]\cos \left(b_{+}+\frac{C}{2v}\right)- \frac{2\kappa v\lambda_{\sigma}}{\gamma^{2}\sqrt{\Delta }\lambda^{2}_{\theta}}\sin b_{+} \cos \frac{C}{2v} \cos \left(b_{+}+\frac{C}{2v}\right)H_{\mathrm{int}},\\
\frac{d \lambda_{\delta}}{d \tau} & = & \frac{v\lambda_{\delta}}{2 \gamma\sqrt{\Delta}}\left[\sin b_{\theta} + \sin \left(b_{+}+\frac{C}{2v}\right)\right]\cos \left(b_{+}-\frac{C}{2v}\right) - \frac{2\kappa v\lambda_{\delta}}{\gamma^{2}\sqrt{\Delta}\lambda^{2}_{\theta}}\sin b_{+} \cos \frac{C}{2v} \cos \left(b_{+}-\frac{C}{2v}\right)H_{\mathrm{int}},\\
\frac{d b_{\theta}}{d \tau} & = & -\frac{v}{\gamma\sqrt{\Delta \nonumber }}\left[2\sin b_{\theta}\sin b_{+}\cos \frac{C}{2v} + \sin \left(b_{+}+\frac{C}{2v}\right)\sin \left(b_{+}-\frac{C}{2v}\right)\right. \\ \nonumber && \left. +(b_{\theta}-b_{+})\left(\sin b_{\theta}\cos b_{+}\cos \frac{C}{2v}+ \frac{1}{2}\sin 2b_{+}\right)\right.
\left. +\frac{C}{2v}\left(\sin b_{\theta}\sin b_{+}\sin \frac{C}{2v}+ \frac{1}{2}\sin \frac{C}{v}\right)\right]\\
&&  + \frac{4\kappa v}{\gamma^{2} \sqrt{\Delta}\lambda^{2}_{\theta}}\sin b_{+}\cos \frac{C}{2v}\left[(b_{\theta}-b_{+})\cos b_{+}\cos \frac{C}{2v}+\frac{C}{2v}\sin b_{+}\sin \frac{C}{2v}\right]H_{\mathrm{int}}+\frac{\kappa\sqrt{\Delta}\lambda_{\theta}^{2}}{8v}H_{0}, \\
\frac{d b_{+}}{d \tau} & = & -\frac{v}{\gamma\sqrt{\Delta \nonumber }}\left[2\sin b_{\theta}\sin b_{+}\cos \frac{C}{2v} + \sin \left(b_{+}+\frac{C}{2v}\right)\sin \left(b_{+}-\frac{C}{2v}\right)\right. \\
&& \left. +(b_{+}-b_{\theta})\cos b_{\theta}\sin b_{+}\cos \frac{C}{2v} +\frac{C}{2v}\left(\sin b_{\theta}\sin b_{+}\sin \frac{C}{2v} +\frac{1}{2} \sin 2b_{-}\right)\right] \nonumber\\
& &  + \frac{4\kappa v}{ \gamma^{2} \sqrt{\Delta } \lambda^{2}_{\theta}}\sin^{2} b_{+}\cos \frac{C}{2v}\left[\cos \frac{C}{2v}+\frac{C}{2v}\sin \frac{C}{2v}\right]H_{\mathrm{int}}, \\
\label{ec:ec_mov_am_2}
\frac{d a_{m}^{(r)}}{d \tau} & = & -i \frac{4v^{2}}{ \gamma^{2} \Delta \lambda^{2}_{\theta}|m|}\sin^2 b_{+}\cos^{2} \frac{C}{2v}\left(a_{m}^{(r)}+a_{-m}^{(r)\ast}\right)-\frac{i}{8}\lambda_{\theta}^{2}|m|a_{m}^{(r)}.
\end{eqnarray}

\end{widetext}


\begin{thebibliography}{50}
\bibitem{note1} Alternatively, if the inhomogeneities are small, one can think of performing a perturbative expansion in them. Naturally, the Bianchi I metric is recovered at zeroth order in this approach, while the higher-order terms are expected to yield small corrections, which in no way could tame the singularities of the background.
\end{thebibliography}
\end{document}